\documentclass{article}

\begin{document}

   \title{Nonlinear and spin effects in two-photon annihilation of a fermion pair
in an intensive laser wave}
\author{Sergey~M.~Sikach
\and \sl Institute of Physics, National Academy of Sciences, \and
\sl F.Skorina Av. 68, Minsk, 220072, Belarus \and \rm e-mail:
sikach@dragon.bas-net.by \and \rm Tel: (375-17) 284-04-29  Fax:
(375-17) 284-08-79 }

\maketitle


\begin{abstract}
The pattern of calculation of amplitudes of a series of processes
in the field of an intensive laser wave, in which two fermions
$(p; p')$ and two real photons $(k_1; k_2)$ participate, is
considered. In relation to one-photon processes, these processes
are of the second order on $\alpha$, if the wave intensity $\xi
\ll 1$ (i.e., actually absorption from the wave only one quantum).
Otherwise, they are competing and essentially nonlinear.
One-photon processes have a number of the important physical
applications. For example, ${\gamma}e$ and ${\gamma}{\gamma}$
colliders work on their basis.

In DSB the calculation is conducted at the level of reaction
amplitudes. It essentially simplifies both the calculation  and
the form of obtained results; those combinations of amplitudes
which describe the spin effects are easy to calculate. And these
effects  are especially essential in nonlinear processes. The
calculations are conducted in covariant form. Besides compactness,
this provides independence of  the frames of reference and energy
modes. The masses of fermions are taken into account precisely and
are not assumed equal each other (heavy leptons, modes of various
decays). The pattern of calculation is given in the channel
indicated in the title. The amplitudes of this channel are also
applicable for calculating of induced decay of orthopositronium
into two photons. And in this approach there is no necessity to
assume the relative momentum equal to zero.
\end{abstract}


\newpage

    The process under consideration is described by $2^5$ amplitudes. Because of
parity conservation, we need to calculate 16 amplitudes (in our
approach, they are described by 8 formulas). Therefore, here we
can present only the pattern of calculation. The necessary
formulas are cited in accordance with monograph \cite{c1} with the
indication paragraph number,where \S 40  ``An electron in the
field of the plane electromagnetic wave''; \S 89 ``An annihilation
of a positronium''; \S 101  ``Radiation of a photon by an electron
in the field of an intensive electromagnetic wave''. Unlike
\cite{c1}, we assume $\gamma_5 = {\it i} \gamma^0  \gamma^1
\gamma^2  \gamma^3$ (opposite sign) and elementary charge $e
> 0$.

    The calculation of amplitudes will be made in the Diagonal Spin Basis (DSB)
introduced by the author in \cite{c5}. A review of the method and
the calculation of number processes analogous to that under
consideration (but with participation only one photon) is given in
\cite{c6}.

    In any reaction, an even number of fermions participate. Therefore,
the amplitude of each reaction can be represented as a combination
of fermion ``sandwiches'' describing the fermion line
\begin {equation}
\displaystyle {\bar u}^{{\sigma}'}(p') Q u^{\sigma}(p) = Tr[ Q
u^{\sigma}(p) {\bar u}^{{\sigma}'}(p')]  \; , \label{e1}
\end {equation}
where $Q$ is the operator of interaction. If it has the tensor
indices, there occurs contraction over them, with other
``sandwiches'' of other fermion lines. In the DSB, the axes of the
spin projections belonging to one fermion line are lying in the
2-plane $(v = p/m, \; v'= p'/ m')$. Vectors $v'$, $s'_3$ and
bispinor $u^{\delta}(p')$ are obtained from $v$, $s_3$ and
$u^{\delta}(p)$ by one and the same plane Lorentz transformation.
\begin{eqnarray}
\displaystyle
\begin{array}{l}
\displaystyle s_3 = { (v v') v - v' \over \sqrt{ (v v')^2 - 1 }}
\; , \; \;
s'_3 = - { (v v') v' - v \over \sqrt{ (v v')^2 - 1 }} \; ;
      \\ \displaystyle
s_1 = s'_1 = n_1 = (\tilde{g} - g){ k \over k_{\perp} } \; , \; \;
s_2 = s'_2 = n_2 = - \tilde{\varepsilon}  { k \over k_{\perp} } \;
;
    \\
\displaystyle  n_0 = {v + v' \over \sqrt{ 2 (v v' + 1) }} \ , \;
 \displaystyle  n_3 = {v - v' \over \sqrt{ 2 (v v' - 1) }} \ ;
\;
\tilde{g}^{\mu \nu} = n_0^{\mu} n_0^{\nu} -  n_3^{\mu} n_3^{\nu} \
, \;
\tilde{\varepsilon}^{\mu \nu} = {{\varepsilon}^{\mu \nu}}_{\rho
\sigma} n_0^{\rho} n_3^{\sigma} \ .
\end{array}
\label{e2}
\end{eqnarray}
 The bispinors of both particles have a common set of spin
operators and are related by the relation
$\displaystyle u^{\delta}(p') = {\hat{n}}_0 u^{\delta}(p) $,
$\displaystyle \delta = \pm 1 $.
If the initial or the final (our case) particle is an antifermion,
this corresponds to the creation or annihilation of a pair. In
this case,
$\displaystyle v^{\delta}(p') = \delta \gamma_5 u^{-\delta}(p') $.
In the DSB, the description of the spin and all the calculations
made are covariant. In the case of annihilation, the spin indices
in bispinors have the meaning of helicity in the system of centre
of masses.

    By virtue of the above properties, the fermion line in DSB
acts as a certain single object, and the basic formulas for
calculating the amplitudes (transition operators) acquire the
elementary form:
\begin {equation}
\displaystyle u^{\delta}(p) {\bar v}^{\delta}(p') = {\gamma_5
\over 4} \left[ ( V^+  + \delta \gamma_5 V^-) (1 - \delta \gamma_5
{\hat{n}}_0 {\hat{n}}_3 ) - {\hat{n}}_0 - \delta \gamma_5
{\hat{n}}_3 \right] \; , \label{e3}
\end {equation}
\begin {equation}
\displaystyle u^{\delta}(p) {\bar v}^{-\delta}(p') = {1 \over 4} (
V^+  + \delta \gamma_5 V^- +  {\hat{n}}_0) ({\hat{n}}_1 + {\it i}
\delta  {\hat{n}}_2 ) \; , \label{e4}
\end {equation}
where
$\displaystyle V^{\pm} = \sqrt{ (v v' \pm 1)/ 2 }$.
As vector $k$, which enters into the definition of $n_1$ and
$n_2$, we take the momentum of laser photon. The replacement of
this vector by another leads to remultiplication of Eq.\ref{e4} by
the easily calculated phase factor.

If in tetrad $n_A$ Eq.\ref{e2} for fermion line we make the
replacement $p \rightarrow k_1$, $p' \rightarrow k_2$, then we
obtain tetrad  $h_A$ common for both photons. In this case, the
vector
$\displaystyle e_{\lambda} = { 1 \over {\sqrt 2}} (h_1 - {\it i}
\lambda h_2)
$
simultaneously describes both the emitted photons: $k_1$  with
helicity  $\lambda_1 = \lambda$ and $k_2$  with $\lambda_2 =-
\lambda$. Thus, if $\lambda_1 + \lambda_2 = 0$, then only one
operator ${\hat{e}}_\lambda$ enters into the amplitude, if
$\lambda_1 + \lambda_2 = \pm 2$, then ${\hat{e}}_\lambda$ and
${\hat{e}}_\lambda^\ast$ enter into the amplitude. This fact also
extremely simplifies both the structure and the calculation of the
reaction amplitudes. Thus,
\begin{eqnarray}
\displaystyle
\begin{array}{l}
\displaystyle {\hat{e}}_\lambda^2 = 0 \; , \; \;
{\hat{e}}_\lambda {\hat{e}}_\lambda^\ast = - 1  + {\gamma_5 \over
k_1 k_2 } k_1^{\mu} {\sigma}_{\mu \nu}  k_2^{\nu}  \; ;
      \\ \displaystyle
{\hat{e}}_\lambda \hat{f} {\hat{e}}_\lambda = 2 e_\lambda f
{\hat{e}}_\lambda \; ; \; \;
{\hat{e}}_\lambda \hat{f} {\hat{e}}_\lambda^\ast ={2 \over k_1
k_2} (f k_2 \omega_{-\lambda} {\hat{k}}_1 + f k_1 \omega_{\lambda}
{\hat{k}}_2) \; .
\end{array}
\label{e5}
\end{eqnarray}
where $f$ is arbitrary vector,
$\displaystyle \omega_\lambda = {1 \over 2} (1 + \lambda
\gamma_5)$.

Let us consider in more detail the case of $\lambda_1 + \lambda_2
= 0$. Using Volkov's solution \cite{c1}-\cite{c3} for the electron
in the field of the plane wave (40.7), (101.3) and above
relations, we obtain the spin structure of the amplitude
\begin {equation}
\displaystyle M = - e_\lambda q \left( {1 \over k_1 q} + {1 \over
k_2 q} \right) {\bar v}^{{\delta}'}(p') \left\{ {\hat{e}}_\lambda
+ {\xi^2 m^2 e_\lambda k \over 2 (kp) (kp')} \hat{k} - e \left( {
{\hat{e}}_\lambda \hat{k} \hat{A} \over 2kp } -  { \hat{A} \hat{k}
{\hat{e}}_\lambda \over 2kp'} \right) \right\} u^{\delta}(p)
\; , \label{e6}
\end {equation}
$\displaystyle A = a ( l_1 \cos \varphi + \mu l_2 \sin \varphi)$
is the 4-potential of the wave (101.2), $\mu$  is the
polarization, $\varphi = k x$, $\xi = ea / m$ is the wave
intensity,
$\displaystyle q^{\mu} = p^{\mu} - {\xi^2 m^2 \over 2kp} k^{\mu}$
is the quasimomentum (101.4). Making series expansion in Bessel
functions  $J_n(z)$ (101.7) and integrating in 4-space, we make
sure that amplitude Eq.\ref{e6} of the reaction $ s k + q + q' =
k_1 + k_2$ completely coincides, to an accuracy of the factor
preceding the ``sandwich'', with the $s$-channel of single-photon
reaction $ s k + q = q' + k'$ (101.10). $s k$ denotes coherent
absorption from the wave of photons (nonlinearity). The factor
itself just reflects that exactly two-photon annihilation takes
place.

For laser wave we choose
\begin {equation}
\displaystyle l_1 = {  (k v')v - (k v) v' \over \sqrt{ (v v')^2 -
1 } k_{\perp} }
\; , \; \; l_2 = n_2 \; . \label{e7}
\end {equation}
Than in (101.6) $\alpha_2 = 0$  and
$\displaystyle z = | \alpha_1 | = { \xi m | kv'/kv - kv/kv'| \over
\sqrt{ (v v')^2 - 1 } k_{\perp} }
$
is the Bessel function argument. With such a choice in the DSB we
have (see Eq.\ref{e4})
\begin{eqnarray}
\displaystyle
\begin{array}{l}
\displaystyle e { {\hat{e}}_\lambda \hat{k} \hat{A} \over 2 k p}
u^{\delta}(p) =
\\ \displaystyle
\mu \delta {\xi {\hat{e}}_\lambda \hat{k} \over 2 m^2  k_{\perp} }
\sum\limits_{s = -\infty}^{\infty} e^{ - {\it i} s \varphi }
\left\{ \left(1 + \mu \delta { s_3 k \over v k } \right) J_{s -
\mu}(z) \omega_{- \mu} - \left(1 - \mu \delta { s_3 k \over v k }
\right) J_{s + \mu}(z) \omega_{\mu}  \right\} u^{\delta}(p)
\; .
\end{array}
\label{e8}
\end{eqnarray}
Besides,
$$\displaystyle {\hat{e}}_\lambda \hat{k} \omega_{-\lambda} =
\sqrt{ { 2 k_2 k \over (k_1 k) (k_1 k_2) } } {\hat{k}}_1 \hat{k}
\omega_{-\lambda} \; , \; \; {\hat{e}}_\lambda \hat{k}
\omega_{\lambda} = \sqrt{ { 2 k_1 k \over (k_2 k) (k_1 k_2) } }
{\hat{k}}_2 \hat{k} \omega_{\lambda} \; . $$
A similar expression is obtained for the structure
$\displaystyle v^{{\delta}'}(p') \hat{A} \hat{k} {\hat{e}}_\lambda
$.
Than rather trivial calculations by formulas ,
Eqs.\ref{e1}-\ref{e8} remain.

   The obtained formulas permit investigating the spin and nonlinear
effects of a number of interesting processes: in the region of
conversion (linear colliders); induced decay of ortopositronium
with coherent absorption (radiation) of an odd number of photons.

 In the $t$-channel (when $e^{\pm} \rightarrow e^{\pm}$), this is
a double Compton back-scattering (the  Compton back-scattering is
a basic process in the operation of $\gamma$-colliders \cite{c7});
$2\gamma$-creation of pairs in the region of conversion (collider
testing), etc. Examples of comprehensive studies of single-photon
processes are given in \cite{c2}-\cite{c4}, \cite{c6}.

It should be need that the inclusion of weak interactions does not
lead to additional complications in calculations and sometimes
even simplifies them. This is connected with the properties of the
projection operators $\omega_\lambda$, namely
$$\displaystyle \omega_{\lambda} \omega_{{\lambda}'}= {1 + \lambda
{\lambda}' \over 2} \omega_{\lambda} \; , \; \; \omega_{\lambda}
\gamma_5 = \lambda \omega_{\lambda} \; . $$

The author wishes to thank A.V.Berestov, A.L.Bondarev,
O.N.Metelitsa.



\end{document}